\documentstyle[preprint,prl,aps]{revtex}

\begin{document}

\draft

\title{Quantum measurement and decoherence}

\author{G. W. Ford\cite{bill} , J. T. Lewis and R. F. O'Connell\cite{bob}}

\address{School of Theoretical Physics\\
Dublin Institute for Advanced Studies\\
10 Burlington Road\\
Dublin 4, Ireland}

\date{\today }

\maketitle

\begin{abstract}
Distribution functions defined in accord with the quantum
theory of
measurement are combined with results obtained from the quantum
Langevin
equation \ to discuss decoherence in quantum Brownian motion.
Closed form
expressions for wave packet spreading and the attenuation of
coherence of a
pair of wave packets are obtained. The results are exact within
the context
of linear passive dissipation. It is shown that, contrary to
widely accepted
current belief, decoherence can occur at high temperature in
the absence of
dissipation. Expressions for the decoherence time with and
without
dissipation are obtained that differ from those appearing in
earlier
discussions.
\end{abstract}


\narrowtext

The problem of decoherence in quantum systems has been of considerable
recent interest. A sampling of earlier theoretical work appears
in Refs. 
\cite{leggett,walls,savage,zurek,venug}, but for a detailed
survey we refer
to a recent book devoted wholly to the subject \cite{giulini}.
In an
introduction to the contents of this book, Joos surveys the
current
situation and, in discussing the mechanism of decoherence,
states that ``
irreversible coupling to the
environment seems to have
become widely accepted (and even quite popular) during the last
decade, not
least through the various contributions by Woljciech Zurek and
his
collaborators
In this present work we come
to quite the
opposite conclusion and in fact show that, in the high
temperature case
considered by Zurek \cite{zurek} and others, decoherence occurs
in a
characteristic time $\tau _{d}$ [ see Eq. (\ref{21}) below]
that is
independent of the Ohmic decay rate $\gamma $ [see Eq. (19)
below] which
characterizes the strength of the coupling to the environment.
Furthermore,
the formulation we use is exact and enables us to show
explicitly that
previous estimates of the decoherence time arise from an
inconsistent
application of long-time asymptotic formulas to obtain a
short-time result.
Apart from its relevance to the question of classical-quantum
correspondence
and the foundations of quantum mechanics, this work, especially
that part
dealing with entangled states, is clearly relevant to the host
of
experiments on decoherence \cite{haroche}, quantum
teleportation \cite
{zeilinger} and quantum information and computation
\cite{bennett,world}.

Much of the discussion of decoherence has been in terms of the simple
problem of a particle moving in one dimension that is placed in
an initial
superposition state (Schr\"{o}dinger ``cat'' state)
corresponding to two
widely separated wave packets. Decoherence is said to occur
when the
long-time interference pattern is destroyed. The key questions
asked are,
first, under what conditions does decoherence occur and,
second, what is the
decoherence time. Previous discussions of the problem have
either used the
Feynman-Vernon influence functional technique \cite{leggett} or
master
equation techniques\cite{walls,savage,zurek} and have been
confined to the
case of Ohmic dissipation. In either case it is assumed that
the initial
state is decoupled from the environment. Here we use a
different formulation
in terms of quantum distribution functions introduced some time
ago by Ford
and Lewis \cite{ford86}. With this formulation we are able to
obtain exact
closed form expressions for the spreading of a wave packet and
for the
attenuation of interference in the two wave packet problem,
which are valid
for all temperatures and for a very general dissipative
environment. In
particular, we avoid the assumption that the particle is
initially decoupled
from the environment; the particle is in equilibrium with the
environment at
the time it is put into the initial state by a measurement. An
important
conclusion is that decoherence can occur for $\gamma
\rightarrow 0$ (absence
of dissipation). A further feature of the Ford-Lewis
formulation is that the
density matrix for the entire system is employed, i.e., one
does not trace
over the environment to obtain a reduced density matrix as
previous
investigations have done.

In the formulation of Ford and Lewis, we regard the particle as part of a
larger system of particle coupled to a reservoir. Initially (or
in the
distant past) the complete system is in equilibrium at
temperature $T$,
described by the normalized density matrix 
\begin{equation}
\rho _{0}=\frac{e^{-H/kT}}{{\rm Tr}\{e^{-H/kT}\}},  \label{1}
\end{equation}
where $H$ is the {\em system} Hamiltonian. To a measurement of
$x$ one
associates a measuring function $\alpha (x_{1})$ ($x_{1}$ is a
c-number)
such that $\left\| \alpha (x-x_{1})\Phi \right\| ^{2}dx_{1}$ is
the
probability that if the system is in a normalized quantum state
$\Phi $ the
instrument will read between $x_{1}$ and $x_{1}+dx_{1}$. An
example is the
``Gaussian slit'' \cite{feynman} for which 
\begin{equation}
\alpha (x_{1})=(2\pi \sigma _{1}^{2})^{-1/4}\exp
\{-\frac{x_{1}^{2}}{4\sigma
_{1}^{2}}\},  \label{2}
\end{equation}
where $\sigma _{1}$ is the experimental width. It follows that
if $x$ is
measured at time $t_{1}$, the probability that the instrument
will read
between $x_{1}$ and $dx_{1}$ is $W(x_{1},t_{1})dx_{1}$, where 
\begin{equation}
W(x_{1},t_{1})={\rm Tr}\{\alpha \lbrack x(t_{1})-x_{1}]\rho
_{0}\alpha
\lbrack x(t_{1})-x_{1}]^{\dag }\}.  \label{3}
\end{equation}
Here $x(t)$ is the Heisenberg operator at time $t$, 
\begin{equation}
x(t)=e^{iHt/\hbar }xe^{-iHt/\hbar }.  \label{4}
\end{equation}
In the same way, if $x$ is measured at time $t_{1}$ and again
at a later
time $t_{2}$ the probability that the first instrument will
read in range $%
dx_{1}$ and the second in $dx_{2}$ is $%
W(x_{1},t_{1};x_{2},t_{2})dx_{1}dx_{2} $. Using an obvious
shorthand
notation, 
\begin{equation}
W(1,2)={\rm Tr}\{\alpha (2)\alpha (1)\rho _{0}\alpha (1)^{\dag
}\alpha
(2)^{\dag }\}.  \label{5}
\end{equation}
Here in $\alpha (j)=\alpha \lbrack x(t_{j})-x_{j}]$ the index
$j$ is meant
to indicate not only the time $t_{j}$ and instrumental position
$x_{j}$, but
also the instrumental parameters such as a width $\sigma _{j}$.
In this way
one can go on to define higher order distribution functions.

The distribution functions can be expressed in terms of the corresponding
characteristic functions, 
\begin{eqnarray}
W(1) &=&\int_{-\infty }^{\infty }\frac{dk_{1}}{2\pi }\xi
(1)e^{-ik_{1}x_{1}},
\nonumber \\
W(1,2) &=&\int_{-\infty }^{\infty }\frac{dk_{1}}{2\pi
}\int_{-\infty
}^{\infty }\frac{dk_{2}}{2\pi }\xi
(1,2)e^{-i(k_{1}x_{1}+k_{2}x_{2})}.
\label{6}
\end{eqnarray}
Now the key formulas needed from Ford and Lewis \cite{ford86}
is that for
quantum Brownian motion these characteristic functions are
given by the
general formulas 
\begin{eqnarray}
\xi (1) &=&\exp \{-\frac{1}{2}k_{1}^{2}\left\langle
x(t_{1})^{2}\right\rangle \}  \nonumber \\
&&\times \int_{-\infty }^{\infty }\frac{dq_{1}}{2\pi
}\tilde{\alpha}(q_{1}-%
\frac{k_{1}}{2})^{\ast }\tilde{\alpha}(q_{1}+\frac{k_{1}}{2}), 
\nonumber \\
\xi (1,2) &=&\exp
\{-\frac{1}{4}\sum_{j=1}^{2}\sum_{l=1}^{2}\left\langle
x(t_{j})x(t_{l})+x(t_{l})x(t_{j})\right\rangle k_{j}k_{l}\} 
\nonumber \\
&&\times \int_{-\infty }^{\infty }\frac{dq_{1}}{2\pi
}\int_{-\infty
}^{\infty }\frac{dq_{2}}{2\pi
}\prod_{j=1}^{2}\tilde{\alpha}(q_{j}-\frac{%
k_{j}}{2})^{\ast }\tilde{\alpha}(q_{j}+\frac{k_{j}}{2}) 
\nonumber \\
&&\times \exp \{q_{1}k_{2}[x(t_{1}),x(t_{2})]\}.  \label{7}
\end{eqnarray}
Here $\tilde{\alpha}$ is the Fourier transform of the function
$\alpha $
describing the measurement, 
\begin{equation}
\tilde{\alpha}(1)=\int_{-\infty }^{\infty }dx_{1}\alpha
(x_{1})e^{-iq_{1}x_{1}}.  \label{8}
\end{equation}
We should remark that in the derivation of these formulas it
was necessary
to assume that the commutator $[x(t_{1}),x(t_{2})]$ is a
c-number. This is
the case for quantum Brownian motion \cite{ford88}.

We first apply these formulas to obtain an expression for the spreading of a
wave packet. That is, at time $t_{1}$ a measurement is made
with an
associated function of the form (\ref{2}) and then at a later
time $t_{2}$ a
second measurement of the same form is made (with index
$1\rightarrow 2$).
The integrals are all standard Gaussian integrals, and we
obtain the results 
\begin{eqnarray}
W(1) &=&\frac{1}{\sqrt{2\pi \sigma ^{2}}}\exp
\{-\frac{x_{1}^{2}}{2\sigma
^{2}}\},  \nonumber \\
W(1,2) &=&\frac{\exp \{-\frac{1}{2(1-\rho
^{2})}(\frac{x_{1}^{2}}{\sigma ^{2}%
}-2\frac{\rho x_{1}x_{2}}{\sigma \tau }+\frac{x_{2}^{2}}{\tau
^{2}})\}}{2\pi
\sigma \tau (1-\rho ^{2})^{1/2}},  \label{9}
\end{eqnarray}
where (note the misprint in the Eq. (7.18) of \cite{ford86})
\begin{eqnarray}
\sigma ^{2} &=&\sigma _{1}^{2}+\left\langle x^{2}\right\rangle
,  \nonumber
\\
\tau ^{2} &=&\sigma
_{2}^{2}-\frac{[x(t_{1}),x(t_{2})]^{2}}{4\sigma _{1}^{2}}%
+\left\langle x^{2}\right\rangle ,  \nonumber \\
2\sigma \rho \tau &=&-s(t_{2}-t_{1})+2\left\langle
x^{2}\right\rangle .
\label{10}
\end{eqnarray}
In this last, $s(t_{2}-t_{1})$ is the mean square displacement 
\begin{equation}
s(t)=\left\langle \{x(t_{1})-x(t_{1}+t)\}^{2}\right\rangle . 
\label{11}
\end{equation}

The distribution $W(1,2)$ is a Gaussian quadratic form, with mean square
width given by 
\begin{eqnarray}
w^{2}(t) &\equiv &\int_{-\infty }^{\infty }dx_{1}\int_{-\infty
}^{\infty
}dx_{2}(x_{1}-x_{2})^{2}W(1,2)  \nonumber \\
&=&\sigma _{1}^{2}-\frac{[x(t_{1}),x(t_{1}+t)]^{2}}{4\sigma
_{1}^{2}}%
+s(t)+\sigma _{2}^{2}.  \label{12}
\end{eqnarray}
This is an exact general formula for the spreading of a
Gaussian wave
packet, expressed in terms of the mean square displacement and
the
nonequal-time commutator. For the special case when the second
measurement
is made with infinite precision ($\sigma _{2}=0$) and for a
free particle
without dissipation and at zero temperature ($s(t)=0$, $%
[x(t_{1}),x(t_{1}+t)]=i\hbar t/m$) this reduces to the familiar
formula of
elementary quantum mechanics \cite{schiff}.

In the case of an unbound (free) particle the mean square displacement $%
\left\langle x^{2}\right\rangle $ diverges. We can obtain a
simple
expression in this limit if we introduce the conditional
probability 
\begin{equation}
P(x_{2}-x_{1},t_{2}-t_{1})=\lim_{\left\langle
x^{2}\right\rangle \rightarrow
\infty }\frac{W(1,2)}{W(1)}.  \label{13}
\end{equation}
Using the expressions (\ref{9}) we find 
\begin{equation}
P(x,t)=\frac{\exp \{-\frac{x^{2}}{2w^{2}(t)}\}}{\sqrt{2\pi
w^{2}(t)}}.
\label{14}
\end{equation}
Thus, the conditional probability is a normal distribution with
variance $%
w^{2}(t)$. \ Here we should refer to the work of Hakim and
Ambegaokar \cite
{hakim}, who use path integral methods for the special case of
a free
particle interacting with an Ohmic bath to obtain an equivalent
expression
for wave packet spreading. (note the misprint in their
expression (38) for
the width)

Next, we consider the case where the initial measurement forms two widely
separated wave packets, which corresponds to the measurement
function 
\begin{equation}
\alpha (1)=\frac{\exp \{-\frac{(x_{1}-d/2)^{2}}{4\sigma
_{1}^{2}}\}+\exp \{-%
\frac{(x_{1}+d/2)^{2}}{4\sigma _{1}^{2}}\}}{[8\pi \sigma
_{1}^{2}(1+e^{-d^{2}/8\sigma _{1}^{2}})^{2}]^{1/4}},  \label{15}
\end{equation}
where $d$ is the separation of the wave packets, the width of
each being $%
\sigma _{1}$ and $x_{1}$ being the center of the wave packet
pair. The
second measurement is then made with a single slit instrument
corresponding
to a function of the form (\ref{2}) (with index $1\rightarrow
2$). Again,
the integrals are all standard Gaussian integrals, and we
obtain results for 
$W(1)$ and $W(1,2)$. Again, there is a considerable
simplification if we
introduce the conditional probability (\ref{13}). We find 
\begin{eqnarray}
P(x,t) &=&\frac{1}{\sqrt{2\pi w^{2}}(1+e^{-d^{2}/8\sigma
_{1}^{2}})} 
\nonumber \\
&&\times \left( \exp \{-\frac{x^{2}+(\sigma _{1}^{2}+s+\sigma
_{2}^{2})\frac{%
d^{2}}{4\sigma _{1}^{2}}}{2w^{2}}\}\cos
\frac{xd[x(t_{1}),x(t_{1}+t)]}{%
4i\sigma _{1}^{2}w^{2}}\right.  \nonumber \\
&&\left. +\frac{1}{2}\exp
\{-\frac{(x-\frac{d}{2})^{2}}{2w^{2}}\}+\frac{1}{2}%
\exp \{-\frac{(x+\frac{d}{2})^{2}}{2w^{2}}\}\right) . 
\label{16}
\end{eqnarray}
This conditional probability is the sum of three contributions,
corresponding to the three terms within the parentheses. The
second and
third clearly correspond to the sum of probabilities of the
form (\ref{14})
from two single slits, while the first term (that involving the
cosine) is
an interference term. It is of interest to study the ratio,
$a(t_{2}-t_{1})$
of the amplitude of the interference term to twice the
geometric mean of the
other two terms, which we will refer to as the attenuation
factor. We find 
\begin{equation}
a(t)=\exp \{-\frac{(s(t)+\sigma _{2}^{2})d^{2}}{8\sigma
_{1}^{2}w^{2}(t)}\}.
\label{17}
\end{equation}
In general, the interest is in the case where the second
measurement is made
with infinite precision, so in the following discussion we set
$\sigma _{2}$
equal to zero.

For quantum Brownian motion, the mean square displacement and the commutator
are given by the formulas 
\begin{eqnarray}
s(t) &=&\frac{2\hbar }{\pi }\int_{0}^{\infty }d\omega {\rm
Im}\{\alpha
(\omega +i0^{+})\}\coth \frac{\hbar \omega }{2kT}(1-\cos \omega
t), 
\nonumber \\
\lbrack x(t_{1}),x(t_{1}+t)] &=&\frac{2i\hbar }{\pi
}\int_{0}^{\infty
}d\omega {\rm Im}\{\alpha (\omega +i0^{+})\}\sin \omega t. 
\label{18}
\end{eqnarray}
where $\alpha $ is the response function. In the so-called
Ohmic case, where
the mean motion is $m\left\langle \ddot{x}\right\rangle +m\gamma
\left\langle \dot{x}\right\rangle =0$, 
\begin{equation}
{\rm Im}\{\alpha (\omega +i0^{+})\}=\frac{\gamma }{m\omega
(\omega
^{2}+\gamma ^{2})}.  \label{19}
\end{equation}

Consider first the case of vanishingly small dissipation ($\gamma
\rightarrow 0$). Then, setting $\sigma _{2}^{2}=0$ and putting
$s=\frac{kT}{m%
}t^{2}$ and $[x(t_{1}),x(t_{1}+t)]=i\frac{\hbar }{m}t$ in
(\ref{12}) and (%
\ref{17}) we see that 
\begin{equation}
a(t)=\exp \{-\frac{d^{2}}{8\sigma
_{1}^{2}+2\bar{\lambda}^{2}+8\frac{m\sigma
_{1}^{4}}{kTt^{2}}}\}\qquad {\rm no\ dissipation},  \label{20}
\end{equation}
where $\bar{\lambda}=\hbar /\sqrt{mkT}$ is the mean thermal de
Broglie
wavelength. Now the interest is always in the case where the
wave packets
are widely separated, $d\gg \sigma _{1}$. From this expression
we see that
for long time, the attenuation factor will be small (i.e.,
there will be
decoherence) if the temperature is sufficiently high that the
mean de
Broglie wavelength is small compared with the spacing, $d\gg
\bar{\lambda}$.
The characteristic time for decoherence to occur will then be 
\begin{equation}
\tau _{d}=\frac{\sigma _{1}^{2}}{\bar{v}d},  \label{21}
\end{equation}
where $\bar{v}=\sqrt{kT/m}$ is the mean thermal velocity. This
decoherence
time is the time for a particle travelling with the mean
thermal velocity to
traverse the slit width multiplied by the ratio of the slit
width to the
slit spacing. Thus we see that we can have {\em decoherence
without
dissipation} (in the sense that $\tau _{d}$ is independent of
$\gamma $,
which characterizes the strength of the coupling to the
environment).

Next we consider the case of Ohmic dissipation at high temperature, where by
high temperature we mean $kT\gg \hbar \gamma $. Then, using
(\ref{19}) we
see from the formulas (\ref{18}) that 
\begin{eqnarray}
s(t) &=&\frac{2kT}{m\gamma }(t-\frac{1-e^{-\gamma t}}{\gamma
}),  \nonumber
\\
\lbrack x(t_{1}),x(t_{1}+t)] &=&\frac{i\hbar }{m\gamma
}(1-e^{-\gamma t}).
\label{22}
\end{eqnarray}
For short times ($\gamma t\ll 1$) these reduce to the those for
the case of
vanishingly small dissipation, for which the decoherence time
is given by (%
\ref{21}). Thus if $\gamma \tau _{d}=\gamma \sigma
_{1}^{2}/\bar{v}d$ is
small (and this will generally be the case at high temperature)
the
decoherence time will be the same as for the case of
vanishingly small
dissipation, given by (\ref{21}).

These exact results are strikingly different from those obtained by previous
investigators \cite{walls,savage,zurek,venug}. It appears that
the
disagreement arises from the fact that others have implicitly
used a long
time ($\tau _{d}\gg \gamma ^{-1}$) approximation to obtain
characteristic
decay times. To see how this comes about, we evaluate the high
temperature
formulas (\ref{22}) for very long times ($\gamma t\gg 1$).
Putting the
result in (\ref{17}) and setting $\sigma _{1}^{2}=0$ we find
$a(t)\sim \exp
\{-d^{2}\gamma t/\bar{\lambda}^{2}\}$. This is an exponential
decaying with
characteristic time $\gamma ^{-1}\bar{\lambda}^{2}/d^{2}$,
which is exactly
twice the expression for the decoherence time obtained by
previous authors 
\cite{savage,zurek,venug}. But we see that it is inconsistent
with the
assumption of long time used to obtain it. At short times, as
we have seen,
we recover the estimate (\ref{21}). One possible reason why
others have
obtain inconsistent results may be due to the fact that they
assumed that
the system and its environment are initially decoupled whereas,
by contrast,
in the formulation we use the particle state is entangled with
the
environment (i.e., in equilibrium) at the time it is put into
the initial
state by a measurement. Putting this point in another way,
previous
discussions have been in terms of the reduced density matrix
but, as pointed
out by Ambegaokar \cite{ambeg}, 
on such short
time scales
the time evolution does not operate on the reduced density
matrix 
alone.''

In conclusion we have seen that the simple and general formulation of
quantum measurement given in \cite{ford86} provides a powerful
method for
discussing quantum stochastic systems. The formalism is in
terms of quantum
distribution functions and, when combined with results obtained
from the
quantum Langevin equation, has enabled us to obtain exact
explicit
expressions for wave packet spreading and the coherence
attenuation factor.
In discussing the latter we have seen that decoherence occurs
at high
temperature with or without dissipation. In either case the
decoherence time
is the same, given by (\ref{21}). At zero temperature,
decoherence occurs
only in the presence of dissipation.

Note added in proof:

In order to understand more clearly the origin of our (no
dissipation)
result (\ref{20}), we recently showed that it may be derived in
a simple
manner solely within the framework of elementary quantum
mechanics and
equilibrium statistical mechanics \cite{ford01a}. In addition,
we have
recently obtained an explicit general solution of the exact
master equation 
\cite{ford01b}. When applied to the situation considered in
previous
discussions, namely a particle at temperature zero suddenly
coupled to a
bath at high temperature, we are led to an expression for the
decoherence
time differing by a factor of 6 from the conventional result.
We see
therefore that the conventional result corresponds to a
particle that is
``warming up'' over a time of order $\gamma ^{-1}$; our result corresponds
to a particle that is initially at the same temperature as the
bath.

\acknowledgments

GWF and RFO'C wish to thank the School of Theoretical Physics, Dublin
Institute for Advanced Studies, for their hospitality.

\end{document}